\documentclass[conference]{IEEEtran}
\IEEEoverridecommandlockouts
\usepackage{cite}
\usepackage{amsmath,amssymb,amsfonts}
\usepackage{algorithmic}
\usepackage{graphicx}
\usepackage{textcomp}
\usepackage[table]{xcolor}
\definecolor{dingred}{HTML}{ea9999}
\definecolor{dingblue}{HTML}{9fc5e8}
\usepackage{textcomp}
\usepackage{subcaption}
\usepackage{hyperref}
\usepackage{booktabs}
\usepackage{multirow}
\usepackage{diagbox}
\usepackage{graphicx}
\usepackage{textgreek}
\usepackage[export]{adjustbox}
\usepackage{pifont}
\usepackage{tikz}
\newcommand{\colorcirc}[2]{%
  \tikz[baseline=(n.base)]\node[circle,draw=black,fill=#1,inner sep=0.6pt,minimum size=10pt,line width=0.55pt](n){\fontsize{5.5}{5.5}\selectfont\textbf{\textcolor{black}{#2}}};%
}
\usepackage{listings}
\usepackage{etoolbox}
\makeatletter
\newcommand{\FPLappearnotice}{%
\textcolor{red}{%
This paper will appear in the 36th International Conference on Field-Programmable Logic and Applications\\[0.3em]
FPL 2026 \textperiodcentered\ September 7--11, 2026 \textperiodcentered\ Ghent, Belgium
}\par\vspace{0.6em}%
}
\patchcmd{\@maketitle}{\centering}{\centering\FPLappearnotice}{}{}
\makeatother
\lstdefinelanguage{XML}{
  basicstyle=\ttfamily\scriptsize,
  morestring=[b]",
  morecomment=[s]{<!--}{-->},
  commentstyle=\color{green!60!black}\itshape,
  stringstyle=\color{violet!80!black},
  identifierstyle=\color{black},
  keywordstyle={[1]\color{blue!80!black}\bfseries},
  morekeywords={[1]sg_pattern,gather,scatter,sg_link_list,sg_link,
                wireconn,interposer_cut,interdie_wire,layer,fixed_layout,fill},
  keywordstyle={[2]\color{red!70!black}},
  morekeywords={[2]name,type,num_conns,from_type,from_switchpoint,side,
                to_type,to_switchpoint,z_offset,x_offset,y_offset,mux,
                seg_type,num,sg_link_name,sg_name,sg_link,
                offset_start,offset_end,x,y},
  breaklines=true,
  breakatwhitespace=false,
  columns=fullflexible,
  keepspaces=true,
  showstringspaces=false,
  frame=single,
  rulecolor=\color{black},
  xleftmargin=2mm,
  xrightmargin=1mm,
}

\def\BibTeX{{\rm B\kern-.05em{\sc i\kern-.025em b}\kern-.08em
    T\kern-.1667em\lower.7ex\hbox{E}\kern-.125emX}}
\begin{document}

\title{Modeling, Optimizing and Exploring Multi-Die FPGA Routing Architectures
}

\IEEEoverridecommandlockouts

\IEEEoverridecommandlockouts

\author{
    \IEEEauthorblockN{
        Amirhossein Poolad\textsuperscript{*\dag}, 
        Soheil Gholami Shahrouz\textsuperscript{*\dag}, 
        Andrew Boutros\textsuperscript{\ddag}, and 
        Vaughn Betz\textsuperscript{\dag}\thanks{\textsuperscript{*}Both authors contributed equally to this work.}
    }
    \IEEEauthorblockA{
        \textsuperscript{\dag}\textit{University of Toronto, Canada} \qquad \textsuperscript{\ddag}\textit{University of Waterloo, Canada} \\
        \{amir.poolad, s.shahrouz\}@mail.utoronto.ca, andrew.boutros@uwaterloo.ca, vaughn@eecg.utoronto.ca
    }
}

\fontsize{9.5pt}{11pt}\selectfont

\maketitle

\begin{abstract}
Die stacking has enabled 2.5D FPGAs by integrating multiple active dice on a passive silicon interposer for improved yield and capacity, and has paved the way for 3D architectures that stack active dice directly atop one another. In these multi-die devices, the unique electrical and physical characteristics of the underlying die-stacking technology impose limitations on inter-die connection density and latency, necessitating a bespoke inter-die routing architecture. However, the absence of accurate and versatile modeling tools has left most questions about how to best design the inter-die routing architecture unanswered.

To address this gap, we enhance the open-source FPGA CAD tool VTR to flexibly specify a wide range of multi-die routing architectures. The placement and routing engines in VPR have also been augmented to improve optimization for both 2.5D and 3D FPGAs. We model a wide variety of inter-die connections in 2.5D/3D FPGAs that use a 7nm process node for active dice and a 45nm process node for the silicon interposer and several die-crossing technologies.
Using this enhanced version of VTR, we conduct a detailed design space exploration of inter-die routing architecture in 2.5/3D FPGAs. We characterize the impact of die-crossing technology, the number of inter-die connections, and the detailed inter-die routing architecture on critical path delay (CPD) and area. We further investigate how varying the inter-die wire length in interposer-based FPGAs influences the trade-off between routability and CPD. Our results show that with suitable inter-die routing architectures and emerging die-stacking technologies, 2.5D and 3D FPGAs can increase capacity without significant routability or delay trade-offs. Specifically, 3D FPGAs achieve up to 14\% wirelength reduction and 6\% CPD improvement over 2D devices, and remain routable even with existing 10\,\textmu m pitch technologies, while 2.5D FPGAs incur only a 2\% wirelength and 4\% CPD overhead at 32\% inter-die connectivity.

\end{abstract}


\section{Introduction}

Applications such as low-latency deep learning acceleration~\cite{boutros2025field}, pre-silicon ASIC emulation~\cite{elsabbagh2023accelerating,perdomo2024makinote}, and data center networking offload~\cite{jeong2025mangoboost,tarafdar2018galapagos} have driven demand for FPGAs with ever-greater logic capacity.
Historically, this demand was met through two approaches: scaling out to large multi-FPGA cluster systems~\cite{samayoa2023survey}, or relying on Moore's law and process technology advances to deliver higher logic density on a single monolithic die~\cite{658762}.
Both approaches, however, face fundamental limitations: multi-FPGA systems suffer from high-latency, low-bandwidth inter-FPGA links that bottleneck application performance~\cite{wang2026tdm}, while achieving good yield for large monolithic dies has become increasingly difficult with advanced process nodes, particularly in the early stages of a new node's life cycle~\cite{feng2022chiplet}. 

\begin{figure}[t!]
    \centering
    \begin{subfigure}[b]{0.48\columnwidth}
        \centering
        \includegraphics[width=\linewidth]{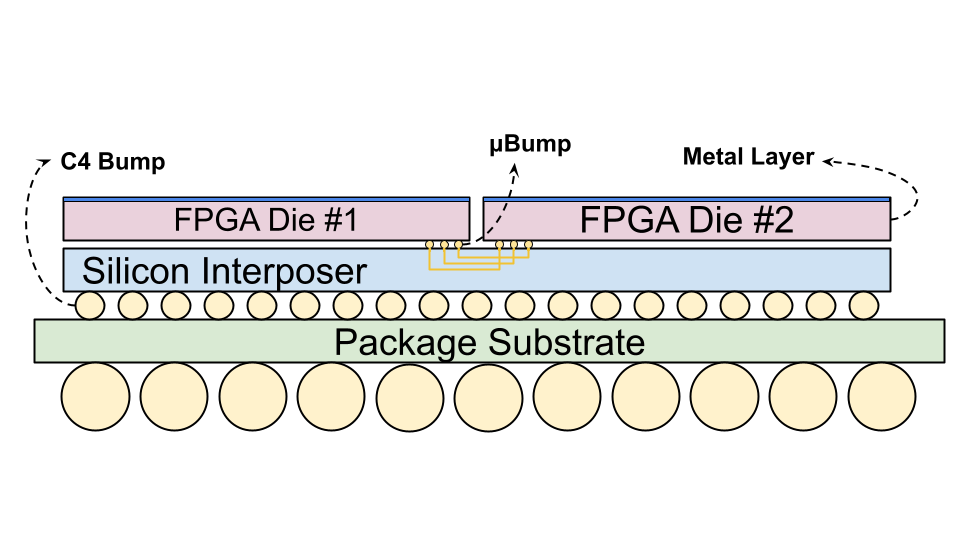}
        \caption{2.5D Integration}
        \label{fig:25d_integration}
    \end{subfigure}
    \begin{subfigure}[b]{0.48\columnwidth}
        \centering
        \includegraphics[width=\linewidth]{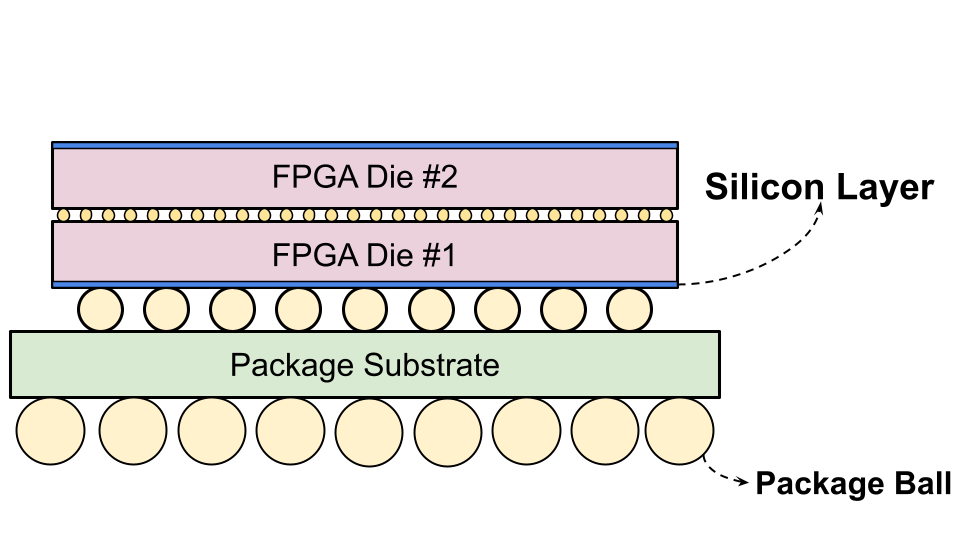}
        \caption{3D Integration}
        \label{fig:3d_integration}
    \end{subfigure}
    \caption{Example of 2.5D and 3D integration in FPGAs.}
    \label{fig:2d_3d_vs}
\end{figure}

Advanced packaging technologies address both limitations by integrating multiple smaller, higher-yield dies into a single package~\cite{li2024high}. Using a passive silicon interposer to connect dies side by side (2.5D integration), or stacking active dies directly atop one another (3D integration), this approach delivers high logic density comparable to monolithic scaling while offering far greater inter-die bandwidth and lower latency than multi-FPGA cluster systems. FPGAs were among the first compute devices to adopt 2.5D integration~\cite{tsmc_xil_press_rel}. Xilinx's Virtex-7~\cite{chaware2012assembly} (28\,nm) and Virtex UltraScale (20\,nm) FPGAs, for example, use passive silicon interposers to integrate three to four FPGA dice; the largest interposer-based devices provide more than twice the logic elements of the largest monolithic FPGA at the same process node~\cite{boutros2021fpga}. Beyond capacity scaling, 2.5D integration also enables heterogeneous devices combining FPGA fabric with memory or processing components in the same package~\cite{versalpaper, agilex_paper}. Several academic works have explored 3D FPGA architectures, demonstrating that stacking active dies vertically yields higher logic density and shorter critical paths~\cite{into_third_dim, youssef2025lazagna, waqar2025monolithic}. Fig.~\ref{fig:2d_3d_vs} illustrates representative 2.5D and 3D FPGA integration schemes.

Inter-die connections introduce fundamental constraints on routing architecture design. In Virtex-7 interposer-based FPGAs, only 23\% of vertical routing tracks cross between dice through the interposer~\cite{tsmc_xil_press_rel}, and each inter-die crossing incurs an additional delay of $\sim$1\,ns~\cite{chaware2012assembly}. In 3D integration, full output pin connectivity across layers requires advanced hybrid-bonding technology, whereas $\mu$bumps limit inter-layer connectivity to only a fraction of output pins~\cite{into_third_dim}. These constraints make routing architecture a particularly critical design consideration in multi-die FPGAs, more so than in conventional 2D devices. Existing studies have explored various aspects of multi-die routing architecture, yet the multifaceted interplay of all relevant parameters has not been fully characterized, leaving room for a more comprehensive treatment. A versatile method to accurately describe and model multi-die routing architectures, capturing parameters such as inter-die connection density, fan-in and fan-out, and for 2.5D devices, the physical length and electrical characteristics of interposer wires, is therefore needed for comprehensive design space exploration.



The main contributions of our work are:
\begin{itemize}
    \item An early-evaluation methodology to assess the feasibility of 2.5D FPGAs across various die stacking technologies.
    \item Detailed electrical modeling of inter-die connections.
    \item A new method to accurately and flexibly describe inter-die routing architectures integrated into the VTR CAD tool's architecture language and architecture generator to enable end-to-end evaluation of 2.5D and 3D FPGAs.
    \item Multi-die aware place and route flow enhancements to best exploit the capabilities of these new systems and ameliorate their weaknesses.
    \item A comprehensive study of 3D and 2.5D routing architectures in terms of delay, area and CAD runtime. Our results show that 3D devices with a better than 2D area-delay product can be built using existing die-stacking technologies that previous studies have ruled out as infeasible. We also show a 25\% variation in area-delay product between our best and worst 2.5D architectures, further indicating the importance of inter-die routing architecture design.
\end{itemize}

\section{Background \& Prior Work}


Semiconductor foundries currently use two main methods for 3D die stacking~\cite{advanced_pack_survey,li2024high}. The first approach uses \textmu-scale solder bumps (\textmu Bumps), in which bumps are deposited on each die and the dice are then soldered together. The second approach, hybrid bonding, eliminates solder entirely by directly fusing copper pads on opposing die surfaces. \textmu Bump-based interconnects are used in the TSMC CoWoS, TSMC SoIC-P, and some iterations of Intel Foveros, with reported pitches of 50\,\textmu m~\cite{foveros_50um}, 45\,\textmu m~\cite{xilinx_first_interposer_paper}, 36\,\textmu m~\cite{foveros_36um}, and 25\,\textmu m~\cite{tsmc_cowos_r}; the International Roadmap for Devices \& Systems (IRDS) projects these pitches reaching as low as 10\,\textmu m in the near future~\cite{irds2024moremoore}. Hybrid bonding is employed in TSMC SoIC-X and Intel Foveros Direct 3D. SoIC-X supports pad pitches from 9\,\textmu m down to 0.3\,\textmu m~\cite{fine_pitch_soic}, with 9\,\textmu m and 6\,\textmu m pitches already in production~\cite{tsmc_25_symposium}, while Intel reports sub-10\,\textmu m pitches for Foveros Direct 3D~\cite{foveros_d3d}. This finer pitch allows hybrid bonding to offer orders of magnitude greater inter-die connectivity than \textmu Bump-based processes. Although hybrid bonding is technically bump-less, throughout this work we use the term "bump" interchangeably for both technologies.

The choice of bump pitch has direct implications for 2.5D FPGA architects, as it governs the trade-off between die-crossing delay and inter-die connection density. The total number of available wires is bounded by the bump count. Extending interposer wires deeper into the die, away from the die edge, exposes more area for bump placement and thereby increases connection density, but comes at the cost of higher wire delay. To quantify this trade-off, Ravishankar et al.~\cite{xil_place_strat} show that for a $7\,\text{mm}\times12\,\text{mm}$ die with a coarse 45\,\textmu m \textmu Bump pitch, wires spanning $\frac{2}{3}$ of the die height, already considerably longer than a standard vertical wire, would amount to only 25\% of the normal vertical routing wires. This illustrates how coarser bump pitches reduce inter-die connection density and increase die-crossing delay, compounding the design challenge. Raikar and Stroobandt~\cite{dirk_how_long} further investigated this trade-off by studying the optimal interposer wire length, finding that longer wires increase the available bump area and inter-die connectivity but at the cost of higher delay.

Beyond the physical constraints imposed by bump pitch, the way interposer wires interface with the FPGA routing fabric also shapes die-crossing performance, and this has evolved across Xilinx architectures. The 7-Series uses a \textit{routing-based} approach, where interposer wires extend the normal routing fabric, with reported delays of around $1$\,ns~\cite{xilinx_25D_3D_pres}. UltraScale adopts a \textit{tile-based} approach, replacing routing resources with specialized tiles featuring an optional register, improving timing at the cost of displacing CLBs. Versal takes a \textit{hybrid} approach, adding an optional interposer connection to each CLB output pin, avoiding CLB displacement while still supporting registered die-crossing paths~\cite{versalpaper}.

These commercial architectures have inspired a body of academic work aimed at modeling interposer-based FPGAs and developing interposer-aware CAD algorithms. Nasiri et al.~\cite{tvlsi16} extend VPR to model routing-based 2.5D devices similar to the 7-Series architecture. They model 2.5D FPGAs as 2D devices with 'interposer cuts' that remove all connections across the cut boundary. Interposer wires are then added to the device to restore inter-die connectivity. Their findings show that an inter-die connectivity ratio of 20\% relative to intra-die connectivity is sufficient for reliable routability without significant delay increase, particularly when interposer wires are bidirectional and have a fan-in and fan-out greater than 1. To mitigate the resulting routability issues and increased delay, they use a simulated annealing-based placer with a cut cost term that seeks to minimize the number of signals that cross between dice, and also evaluate an optional graph partitioning step to further minimize inter-die connections~\cite{tvlsi16}.

Expanding that work, Iyer et al.~\cite{ucsd_part_interposer} built a bespoke timing-aware partitioning tool for 2.5D FPGAs. Ravishankar et al.~\cite{xil_place_strat} introduced a partitioning-based placer targeting AMD's UltraScale architecture. Similarly, Raikar and Stroobandt introduced LiquidMD~\cite{liquidmd}, extending the analytical placement flow Liquid~\cite{liquid} that incorporates a graph partitioning pre-processing step for 2.5D devices. Conversely, Di et al.~\cite{leaps} introduce LEAPS, a purely analytical placer that does not rely on a graph partitioner. While these placement algorithms help reduce die-crossing costs, inter-die latency continues to impact design speed, motivating CAD flows that incorporate interposer-awareness earlier in the design process. To this end, Guo et al.~\cite{autobridge} introduce AutoBridge, a multi-die aware HLS framework that distributes logic across dice and inserts additional pipeline registers into inter-die paths.

The works above focus on 2.5D FPGAs, where dice lie side by side on a passive interposer. A complementary body of research investigates truly 3D FPGAs, where active dice are stacked directly on top of one another. Early work by Gayasen et al.~\cite{gayasen2008designing} and Razavi et al.~\cite{sabzi} examines how the switch block must be redesigned for 3D integration. Boutros et al.~\cite{into_third_dim} propose a 3D FPGA architecture targeting a 7\,nm technology node. In their design, inter-layer connectivity is realized by connecting the output pin of each logic block to a bump that drives intra-die wires on the adjacent layer. Youssef et al.~\cite{youssef2025lazagna} introduce LaZagna, an open-source framework that extends VTR and OpenFPGA~\cite{tang2019openfpga} to model 3D FPGAs. LaZagna supports inter-layer connectivity through both connection blocks and switch blocks. For switch-block-based inter-layer connections, the authors propose a syntax similar to VPR's existing custom switch block pattern specification. Their experiments show that 3D architectures with as few as 26 inter-layer connections per logic block remain routable. Cherukuri et al.~\cite{cherukuri2026vfpga} propose vFPGA, a 3D FPGA in which SRAM configuration bits are stored on a dedicated layer separate from the logic fabric. This decoupling allows the configuration layer to be implemented in an older technology node, which is advantageous because SRAM scaling has slowed considerably at recent process nodes.

\section{Inter-die Connection Physical Modeling}



\subsection{Interposer Feasibility Study}
\label{sec:int_feas}

Unlike 3D FPGAs, which stack dice to allow vertical connections across their entire surface, 2.5D FPGAs place dice side-by-side, restricting connections to the die boundaries. Because interposer wires have a finite reach (e.g., 2 mm), any bumps placed beyond that distance from the edge are useless for inter-die communication. Consequently, 2.5D inter-die connectivity depends on both bump pitch and interposer wire length, introducing an additional dimension in the design space. This motivates a first-order feasibility study to prune the 2.5D design space before undertaking detailed evaluation.

Consider a hypothetical FPGA with square LB tiles of height $H_{LB}$, routing channel width $W$, and bump pitch $P$. Each LB can accommodate at most $n_b = H_{LB}^2/P^2$ bumps in its footprint. To achieve an inter-die connectivity ratio $\alpha$ (i.e., $\alpha \cdot W$ connections cross the die boundary per LB column), the bumps from $\alpha \cdot W / n_b$ rows of LBs along the boundary must be used, requiring an interposer wire length of $L_{int} = \alpha \cdot W \cdot P^2 / H_{LB}^2$. The maximum achievable inter-die connectivity ratio is therefore:
\begin{equation}
    \alpha = \frac{L_{int} \cdot H_{LB}^2}{W \cdot P^2}
    \label{eq:alpha}
\end{equation}
Equation~\ref{eq:alpha} confirms the earlier qualitative argument: $\alpha$ increases with $L_{int}$ and decreases with $P$, and both parameters must be considered when evaluating inter-die connectivity in interposer-based FPGA architectures.

\begin{table}[t]
    \centering
    \footnotesize
    \setlength{\tabcolsep}{3pt}
    \resizebox{0.7\columnwidth}{!}{%
    \begin{tabular}{|c|c|c|c|c|c|c|}
    \hline
    \multicolumn{1}{|c|}{\diagbox{$\alpha$}{$P$}} & \textbf{45 \textmu m} & \textbf{36 \textmu m} & \textbf{25 \textmu m} & \textbf{10 \textmu m} & \textbf{5 \textmu m} & \textbf{1 \textmu m} \\ \hline
    0.1 & \cellcolor[HTML]{EEFFBF}101 & \cellcolor[HTML]{C1FFBF}75 & \cellcolor[HTML]{BFFFBF}30 & \cellcolor[HTML]{BFFFBF}10 & \cellcolor{green!25}1.25 & \cellcolor{green!25}0.05 \\ \hline
    0.2 & \cellcolor[HTML]{FFBFBF}202 & \cellcolor[HTML]{FFBFBF}150 & \cellcolor[HTML]{BFFFBF}60 & \cellcolor[HTML]{BFFFBF}20 & \cellcolor{green!25}2.5 & \cellcolor{green!25}0.10 \\ \hline
    0.3 & \cellcolor{red!25}302 & \cellcolor[HTML]{FFBFBF}225 & \cellcolor[HTML]{DDFFBF}90 & \cellcolor[HTML]{BFFFBF}30 & \cellcolor{green!25}3.75 & \cellcolor{green!25}0.15 \\ \hline
    0.4 & \cellcolor{red!25}403 & \cellcolor[HTML]{FFBFBF}300 & \cellcolor[HTML]{FFE9BF}120 & \cellcolor[HTML]{BFFFBF}40 & \cellcolor{green!25}5 & \cellcolor{green!25}0.20 \\ \hline
    0.5 & \cellcolor{red!25}503 & \cellcolor{red!25}375 & \cellcolor[HTML]{FFBFBF}150 & \cellcolor[HTML]{BFFFBF}50 & \cellcolor{green!25}6.25 & \cellcolor{green!25}0.25 \\ \hline
    0.6 & \cellcolor{red!25}604 & \cellcolor{red!25}450 & \cellcolor[HTML]{FFBFBF}180 & \cellcolor[HTML]{BFFFBF}60 & \cellcolor{green!25}7.5 & \cellcolor{green!25}0.30 \\ \hline
    0.7 & \cellcolor{red!25}705 & \cellcolor{red!25}525 & \cellcolor[HTML]{FFBFBF}210 & \cellcolor[HTML]{C1FFBF}70 & \cellcolor{green!25}8.75 & \cellcolor{green!25}0.35 \\ \hline
    0.8 & \cellcolor{red!25}805 & \cellcolor{red!25}600 & \cellcolor[HTML]{FFBFBF}240 & \cellcolor[HTML]{CEFFBF}80 & \cellcolor{green!25}10 & \cellcolor{green!25}0.40 \\ \hline
    0.9 & \cellcolor{red!25}906 & \cellcolor{red!25}675 & \cellcolor[HTML]{FFBFBF}270 & \cellcolor[HTML]{DDFFBF}90 & \cellcolor{green!25}11.25 & \cellcolor{green!25}0.45 \\ \hline
    1 & \cellcolor{red!25}1006 & \cellcolor{red!25}750 & \cellcolor[HTML]{FFBFBF}300 & \cellcolor[HTML]{EEFFBF}100 & \cellcolor{green!25}12.5 & \cellcolor{green!25}0.50 \\ \hline
    \end{tabular}
    }
    \caption{Interposer wire length ($\times H_{LB}$) required to achieve inter-die connectivity fraction~$\alpha$ for different pitches. Cells are color-coded by RC delay, from lowest (green) to highest (red).}
    \label{tab:int_wl_vs_crossing}
\end{table}

In this work, we use the 7\.nm FPGA fabric introduced in~\cite{into_third_dim}, with $H_{LB} = 24.6\,\mu\text{m}$, and a channel width of $W = 300$. Table~\ref{tab:int_wl_vs_crossing} shows the interposer wire length (in units of $H_{LB}$) required to achieve various inter-die connectivity ratios $\alpha$ for different pitch sizes, as given by Eq.~\ref{eq:alpha}.

\begin{figure}
    \centering
    \includegraphics[width=0.9\linewidth]{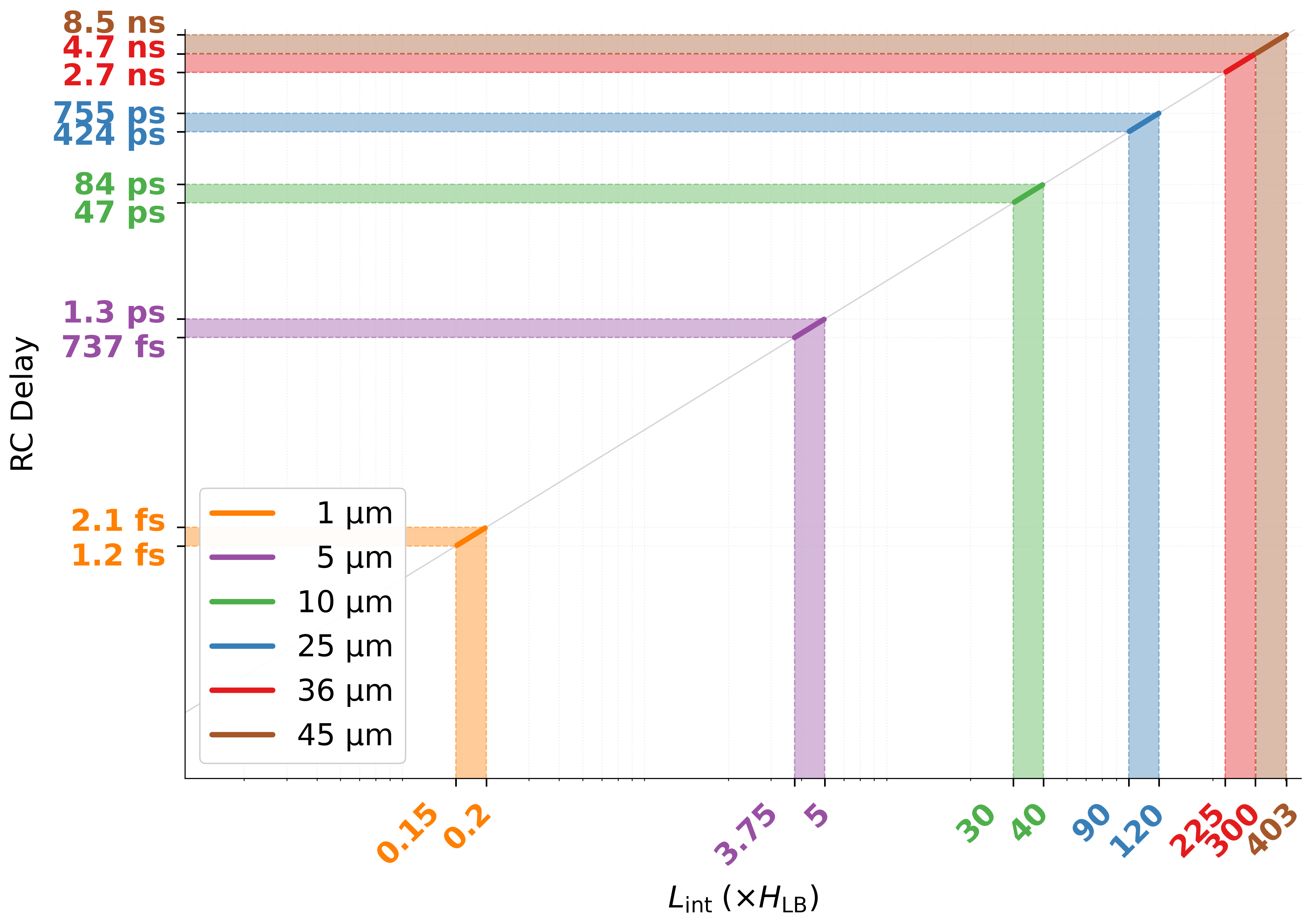}
    \caption{Distributed RC delay of interposer wires vs. their length, assuming a 45\.nm process node. Highlighted regions indicate the wire length ranges and associated delay ranges required to achieve inter-die connectivity ratios of $\alpha \in [0.3,\, 0.4]$.}
    \label{fig:int_rcdelay_model}
\end{figure}

To estimate the intrinsic delay of interposer wires, we assume the interposer uses a 45\.nm technology node with inter-die wires implemented on the global metal layers. The electrical and physical characteristics of the global metal layer are extracted from~\cite{ITRS2005Interconnect}. Using a distributed RC delay model, Fig.~\ref{fig:int_rcdelay_model} plots intrinsic delay against wire length. The highlighted regions indicate the wire length ranges and their associated delay ranges needed to achieve $\alpha \in [0.3,\, 0.4]$ for different pitch sizes. This range is chosen because early experiments indicated it is the minimum inter-die connectivity ratio that yields reliably routable FPGA devices.

For $P = 45-36\,\mu\text{m}$, achieving $\alpha \in [0.3,\, 0.4]$ requires interposer wires spanning hundreds of LBs, with an estimated intrinsic wire delay of 2.7--8.5\,ns. A delay of this magnitude imposes a significant penalty on the critical path delay and makes high inter-die connectivity extremely costly. Furthermore, the required interposer wire lengths are extremely long, meaning that each wire crossing the die boundary must span hundreds of LBs. This can confuse placement and routing algorithms, as tiles physically adjacent across the inter-die boundary may not be reachable through short local routing wires. For $P = 25\,\mu\text{m}$, the required wire lengths and associated delays are substantially reduced compared to the larger pitches, with intrinsic RC delays in the range of hundreds of picoseconds, but the needed interposer wire lengths remain very long.
At $P = 10\,\mu\text{m}$, a reasonable inter-die connectivity ratio can be achieved with interposer wires spanning only tens of LBs, comparable to intra-die long wires, making this pitch a practical candidate. With $P = 5\,\mu\text{m}$ and $P = 1\,\mu\text{m}$, $\alpha = 1$ can be reached with short interposer wires and negligible intrinsic delay, suggesting that near-full inter-die connectivity is physically feasible. However, very short interposer wires introduce a different concern: since inter-die routing resources are accessed via intra-die routing wires, concentrating a large number of inter-die connections near the die boundary can cause routing congestion in that region.

The existence of the Xilinx Virtex-7 family, a 2.5D FPGA that uses a bump pitch of $45\,\mu\text{m}$, may appear to contradict our analysis. This discrepancy is explained by several differences between the 7\.nm device we analyze and the 28\.nm Virtex-7. Most importantly, the Virtex-7 LBs are much larger than those of our 7\,nm fabric (we would expect their $H_{LB}$ to be approximately 4$\times$ as large) and Equation~\ref{eq:alpha} shows that $\alpha$ has a quadratic dependence on $H_{LB}$. Second, Virtex-7 uses only approximately 50 die-crossing wires per routing channel rather than the $[90,\, 120]$ implied by $\alpha \in [0.3,\, 0.4]$ with $W = 300$; applying our model at 50 connections yields an estimated RC delay of approximately 1.5\,ns, which is in the range of the 1\.ns interposer delay of Virtex-7~\cite{tvlsi16}. Finally, our estimates assume minimum-pitch interposer wires; widening the wires would reduce resistance and improve delay.




\subsection{Detailed Interconnect Modeling}\label{sec:detailed_modeling}

\begin{figure}[t]
    \centering
    \begin{subfigure}{\columnwidth}
        \centering
        \includegraphics[width=0.34\linewidth]{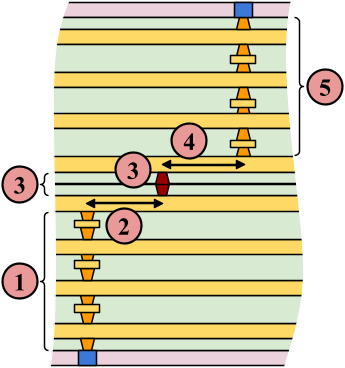}
        \includegraphics[width=0.6\linewidth]{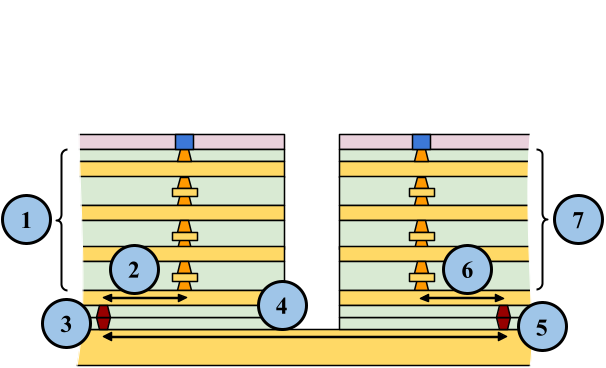}
        \label{fig:3d_model}
    \end{subfigure}

    \vspace{1em} 

    \begin{subfigure}{\columnwidth}
        \centering
        \includegraphics[width=1\columnwidth, valign=c]{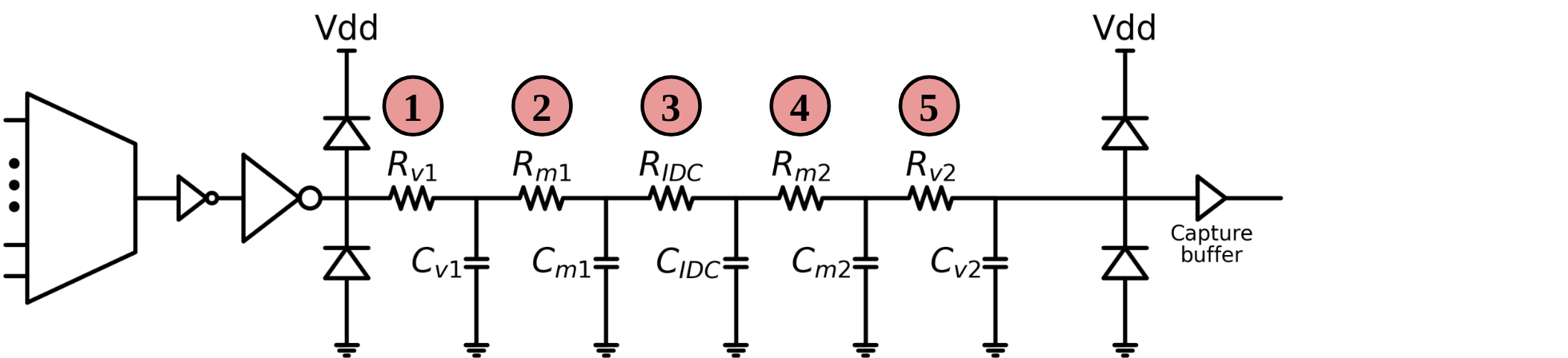}
        \hfill
        \includegraphics[width=1\columnwidth]{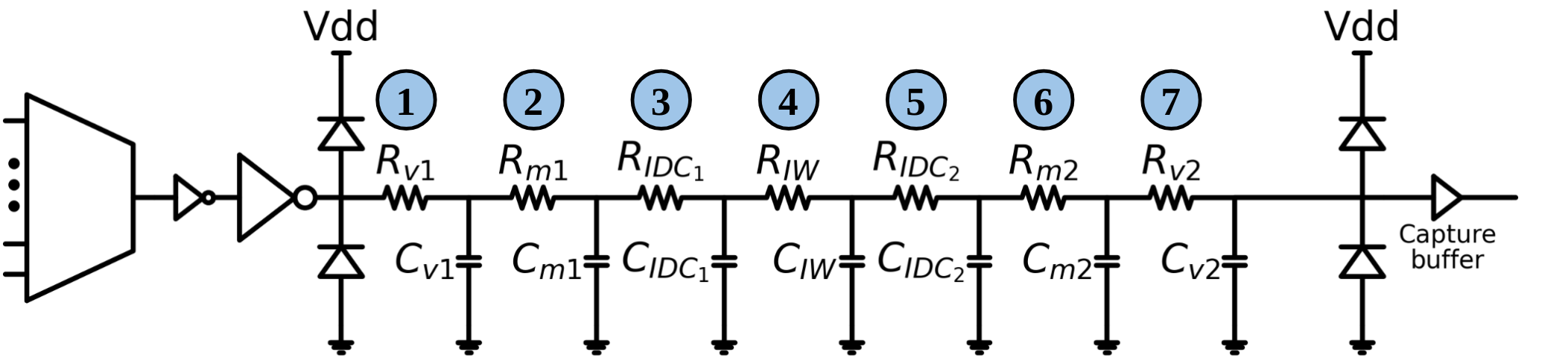}
        \label{fig:25d_model}
    \end{subfigure}
    \caption{2.5D (blue) and 3D (red) inter-die connection circuit models used for detailed interconnect analysis.}
    \label{fig:i2d_conn_model}
\end{figure}

The wire delays reported in Fig.~\ref{fig:int_rcdelay_model} are intrinsic RC delays of the interposer wire alone and do not account for drivers, multiplexers, or ESD protection circuitry. For a rigorous exploration of inter-die routing architectures, we need accurate delays based on realistic circuit models. To this end, we perform SPICE simulations of 2.5D and 3D interconnect models to obtain accurate delay and area estimates. The 3D model is derived from~\cite{into_third_dim}, while the 2.5D model is an extension we introduce in this work. Fig.~\ref{fig:i2d_conn_model} shows both circuit models. Each model consists of an input multiplexer, a two-stage driver, a series of distributed RC loads, and ESD protection circuitry~\cite{rosenbaum2012esd}. In the 3D model, the RC loads represent the signal path from \colorcirc{dingred}{1}~the source die's silicon layer and through its via stack, \colorcirc{dingred}{2}~a worst-case distance of half the perimeter of an SRAM macro on top-level metal to reach the inter-die connection, \colorcirc{dingred}{3}~the inter-die connection, \colorcirc{dingred}{4}~the destination die's top-level metal, and \colorcirc{dingred}{5}~the destination die's via stack. The 2.5D model extends the 3D model by inserting an interposer wire between the two bumps: after passing through \colorcirc{dingblue}{1}~the source die's via stack and \colorcirc{dingblue}{2}~its top-level metal to reach \colorcirc{dingblue}{3}~the first bump, the signal traverses \colorcirc{dingblue}{4}~the interposer wire, \colorcirc{dingblue}{5}~the second bump, \colorcirc{dingblue}{6}~the destination die's top-level metal, and finally \colorcirc{dingblue}{7}~the destination die's via stack. Interposer wire resistance and capacitance are taken from the calculations in Section~\ref{sec:int_feas}, while the active-die metal and via stack parameters are based on a 7\,nm node process~\cite{clark2016asap7, nikolic2021global}, and the inter-die connection and bump parasitics are adopted from~\cite{zhu2022power}.

\begin{figure}[t]
    \centering
    \includegraphics[width=\linewidth]{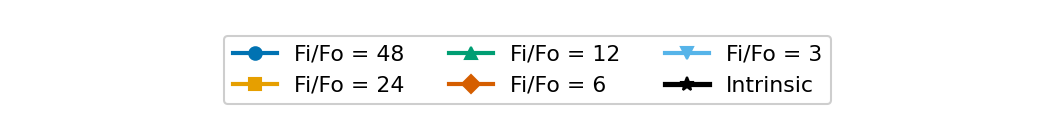}\\[-0.6ex]
    \begin{subfigure}[b]{0.48\columnwidth}
        \centering
        \includegraphics[width=\linewidth]{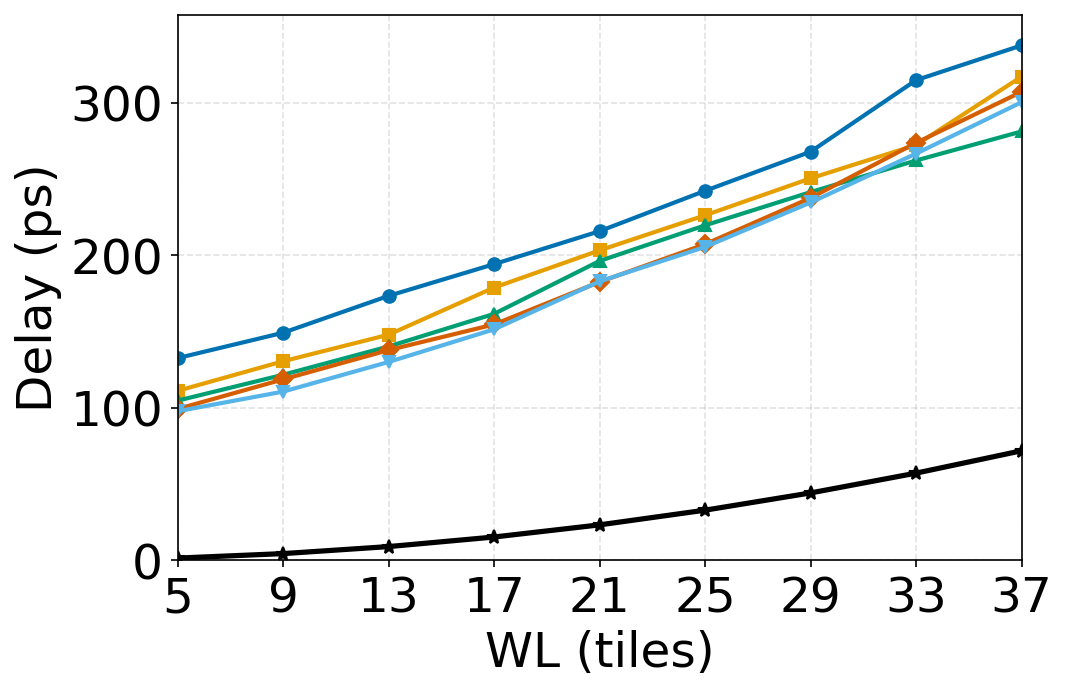}
        \caption{$10\,\mu\text{m}$ pitch.}
        \label{fig:delay_vs_wl_10um}
    \end{subfigure}
    \hfill
    \begin{subfigure}[b]{0.48\columnwidth}
        \centering
        \includegraphics[width=\linewidth]{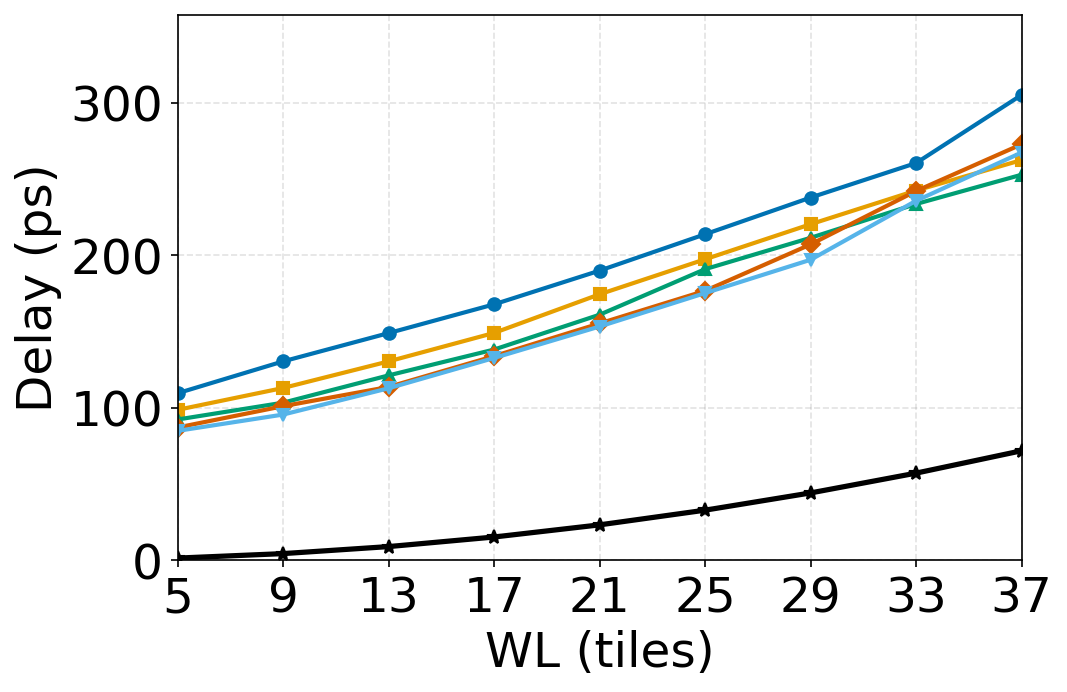}
        \caption{$5\,\mu\text{m}$ pitch.}
        \label{fig:delay_vs_wl_5um}
    \end{subfigure}
    \caption{Total interconnect delay vs.\ wirelength for varying fan-in/fan-out, with intrinsic Elmore delay reference.}
    \label{fig:delay_vs_wl_pitch}
\end{figure}

For each combination of die-crossing technology, input multiplexer size, and interposer wire length (for 2.5D models), we sweep over a range of driver stage ratios and gather mux transistor sizes in HSPICE and select the configuration that minimizes the $\text{area} \times \text{delay}^2$ product. Delay and area are similarly simulated for 3D inter-die connections, though these results are omitted here for brevity. Fig.~\ref{fig:delay_vs_wl_pitch} plots the simulated total delay versus interposer wire length for the 2.5D case. Compared to the intrinsic RC delay of Section~\ref{sec:int_feas}, the full circuit delay is substantially higher due to multiplexer, driver, and ESD overhead. Higher fan-in/fan-out increases delay through higher mux and output capacitance. The $10\,\mu\text{m}$ pitch shows slightly higher delay than $5\,\mu\text{m}$ because of the switch from \textmu Bumps to the bump-less hybrid bonding technology and the subsequent reduction in resistance and capacitance.

%

\section{Architectural Modeling and CAD Flow for Multi-die FPGAs}

\subsection{Scatter-Gather Syntax}\label{sec:sg_syntax}

Having characterized the delay and area of inter-die connections, these parameters must now be incorporated into the CAD flow. To flexibly specify inter-die routing architectures in VTR, we introduce the \textit{scatter-gather} (SG) syntax as an extension to the VTR architecture description XML format~\cite{luu2011architecture}. The SG abstraction models an inter-die connection, illustrated in Fig.~\ref{fig:sg_syntax}: a \textit{gather} phase aggregates $F_g$ local routing wires from a source switch block through a multiplexer, and a \textit{scatter} phase fans out the multiplexed signal to $F_{sc}$ wires at a destination switch block, which may reside on a different die or a distant location on the same die.

\begin{figure}[t]
\centering
{\setlength{\fboxrule}{0.3pt}
\begin{lstlisting}[language=XML, basicstyle=\ttfamily\scriptsize]
<sg_pattern name="2.5D_conn" type="unidir">
 <!-- Gather fan-in=3 from bottom, left, right -->
 <gather>
  <wireconn num_conns="3" from_type="L4"
            from_switchpoint="0" side="blr"/>
 </gather>
 <!-- Scatter fan-out=4 to all sides -->
 <scatter>
  <wireconn num_conns="4" to_type="L4"
            to_switchpoint="0" side="rltb"/>
 </scatter>
 <sg_link_list>
  <!-- Cross interposer cut (y_offset=+3, L_int=3) -->
  <sg_link name="UP" y_offset="3" mux="2.5D_SB_MUX"
           seg_type="int_wire"/>
 </sg_link_list>
</sg_pattern>

<!-- Horizontal cut; two equally sized dies -->
<interposer_cut y="H/2">
 <!-- Instantiate N_sg=2 UP links -->
 <interdie_wire sg_name="2.5D_conn" sg_link="UP" num="2"
               offset_start="-1" offset_end="-2"/>
</interposer_cut>
\end{lstlisting}
}
\caption{SG XML syntax for 2.5D interposer-based FPGAs.}
\label{fig:sg_syntax}
\end{figure}

An SG pattern is specified via the \texttt{<sg\_pattern>} tag, which supports two directionality modes: \texttt{unidir} (the inter-die wire is driven by a single non-tristatable driver from one end) and \texttt{bidir} (the inter-die wire can be driven by a tri-statable driver at either end). The \texttt{<gather>} tag contains one or more \texttt{<wireconn>} elements, each selecting a wire type from a source switch block, with \texttt{num\_conns} setting the fan-in $F_g$ and \texttt{side} restricting which switch-block sides contribute. The \texttt{<sg\_link>} tag specifies the multiplexer type, the inter-die wire segment, and the spatial displacement to the destination switch block via spatial offsets. The \texttt{<scatter>} tag mirrors the gather, distributing the signal to $F_{sc}$ wires at the destination switch block.

Fig.~\ref{fig:sg_syntax} shows an example SG pattern for a 2.5D interposer-based architecture, where the \texttt{y\_offset} on each \texttt{<sg\_link>} encodes the interposer wire displacement and length $L_{int}$. For brevity, only the upward link (\texttt{y\_offset=+3}) is shown; a downward link can be instantiated identically. We additionally introduce the \texttt{<interposer\_cut>} tag, which specifies the inter-die boundary location across which there is a discontinuity in the routing fabric and instantiates SG patterns in a user-defined area around it to model inter-die connectivity. Fig.~\ref{fig:25d_routing_arch} illustrates the resulting routing architecture. The same syntax naturally extends to 3D stacked architectures by using \texttt{z\_offset} instead of \texttt{y\_offset} or \texttt{x\_offset}. Together, the parameters $N_{sg}$ (connections per logic block), $F_g$/$F_{sc}$ (fan-in/fan-out), and $L_{int}$ (inter-die wire length) fully parameterize inter-die routing architecture design for both 2.5D and 3D FPGAs.

\subsection{Multi-die Aware FPGA CAD Flow}\label{sec:cad_flow}
\subsubsection{Routing Resource Graph Construction}

Each SG connection is realized in the routing resource (RR) graph as a new node with $F_g$ incoming edges from local routing wires at the source switch block and $F_{sc}$ outgoing edges fanning out at the destination. SG nodes are appended as a post-processing step, keeping the implementation modular and independent of base graph construction.

For 3D stacked architectures, we introduce a new RR node type, \texttt{CHANZ}, characterized by a \emph{layer range} encoding the range of stacked layers it connects. This is more explicit and expressive than the approaches taken by prior work. Boutros et al.~\cite{into_third_dim} connected \texttt{OPIN} nodes directly to \texttt{CHANX}/\texttt{CHANY} nodes on the adjacent layer, limiting inter-layer connections to those driven solely by output pins and ruling out bidirectional links. VTR 9 introduced 3D switch blocks and models inter-layer connections as two \texttt{CHANX} nodes joined by a single edge~\cite{vtr9}. While this allows switch block connections between layers, it misleads the router lookahead into treating vertical die crossings as horizontal displacements, and the 3D switch block specification does not allow detailed control over the multiplexing to and from the scarce 3D interconnect. Lazagna~\cite{youssef2025lazagna} augments the VTR 9 code base with an external RR graph generator to insert 3D connections, and hence inherits the same restrictions. The \texttt{CHANZ} node, combined with the new SG syntax, eliminates all of these limitations.

In interposer-based 2.5D FPGAs, intra-die wires must not cross the die boundary defined by the \texttt{<interposer\_cut>} tag. Prior works used an RR graph post-processing step to detect and shorten straddling wires~\cite{tvlsi16}, reducing connectivity near the boundary. Our approach instead enforces the constraint during RR graph generation: any wire that would cross the \texttt{<interposer\_cut>} boundary is split into two shorter wires, one on each side.

\subsubsection{Router Lookahead}

\begin{figure}[t]
     \centering
     \begin{subfigure}[b]{0.32\columnwidth}
         \centering
         \includegraphics[width=\textwidth]{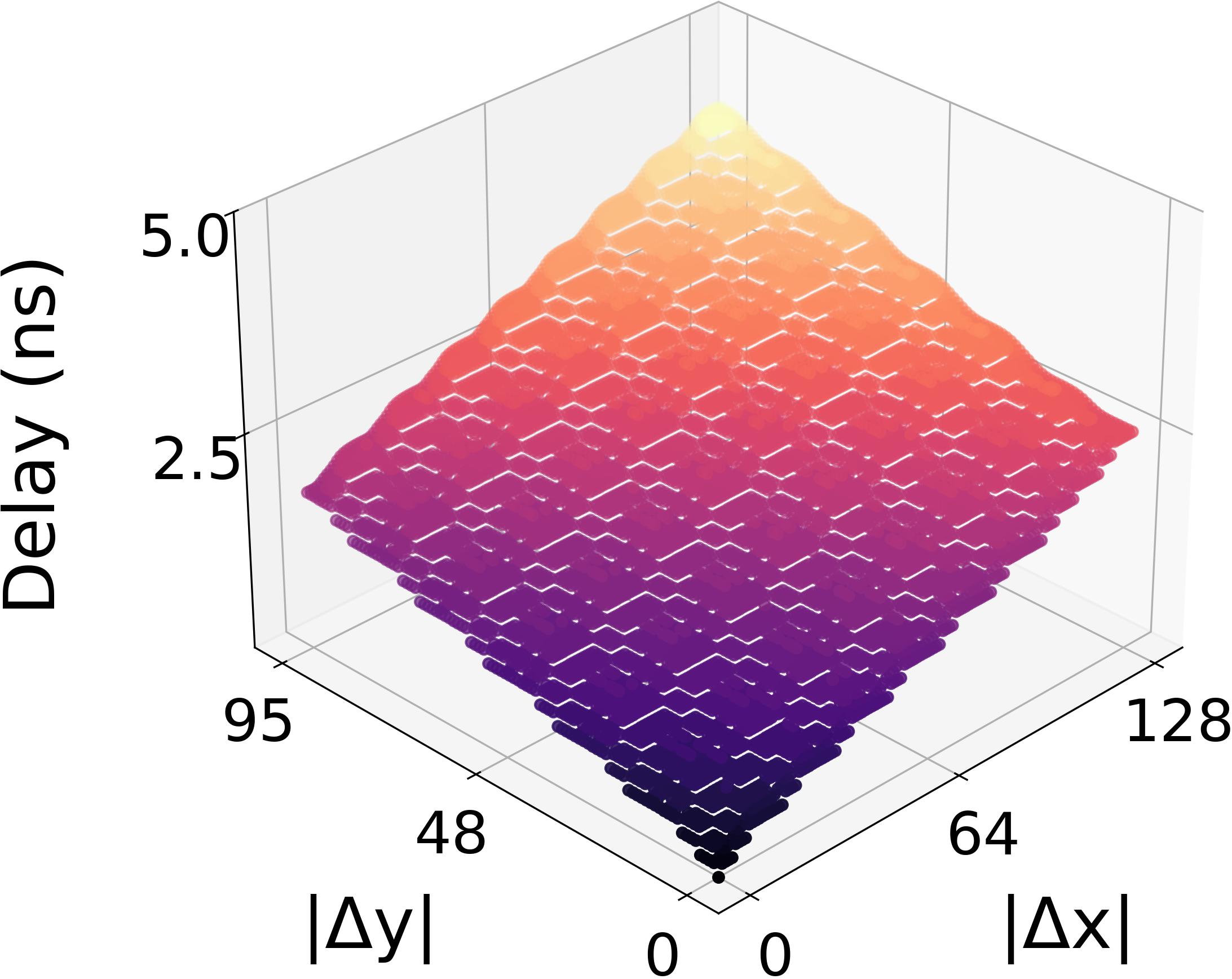}
         \caption{}
         \label{fig:lookahead_no_int}
     \end{subfigure}
     \hfill
     \begin{subfigure}[b]{0.32\columnwidth}
         \centering
         \includegraphics[width=\textwidth]{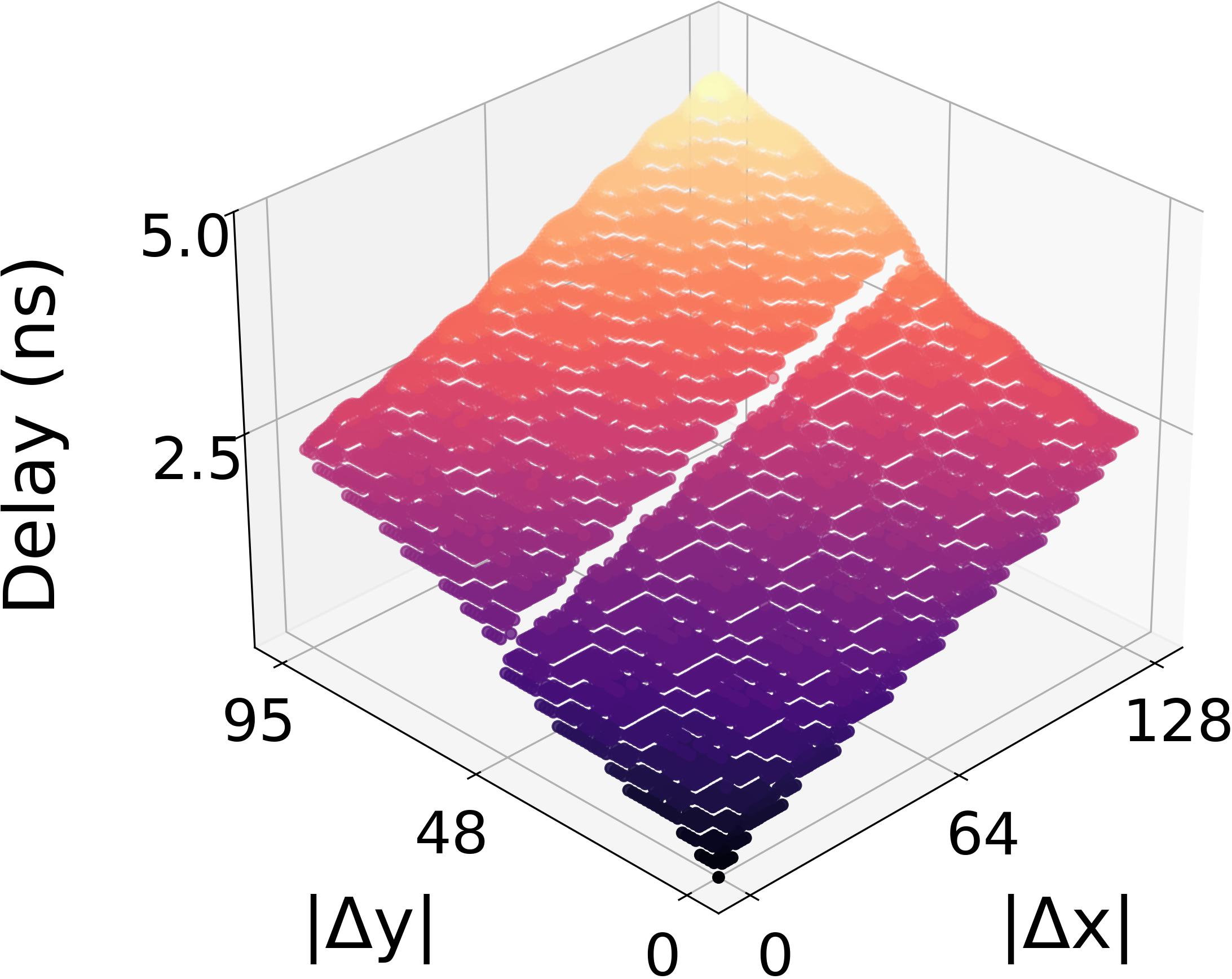}
         \caption{}
         \label{fig:fig:lookahead_int}
     \end{subfigure}
     \hfill
     \begin{subfigure}[b]{0.32\columnwidth}
         \centering
         \includegraphics[width=\textwidth]{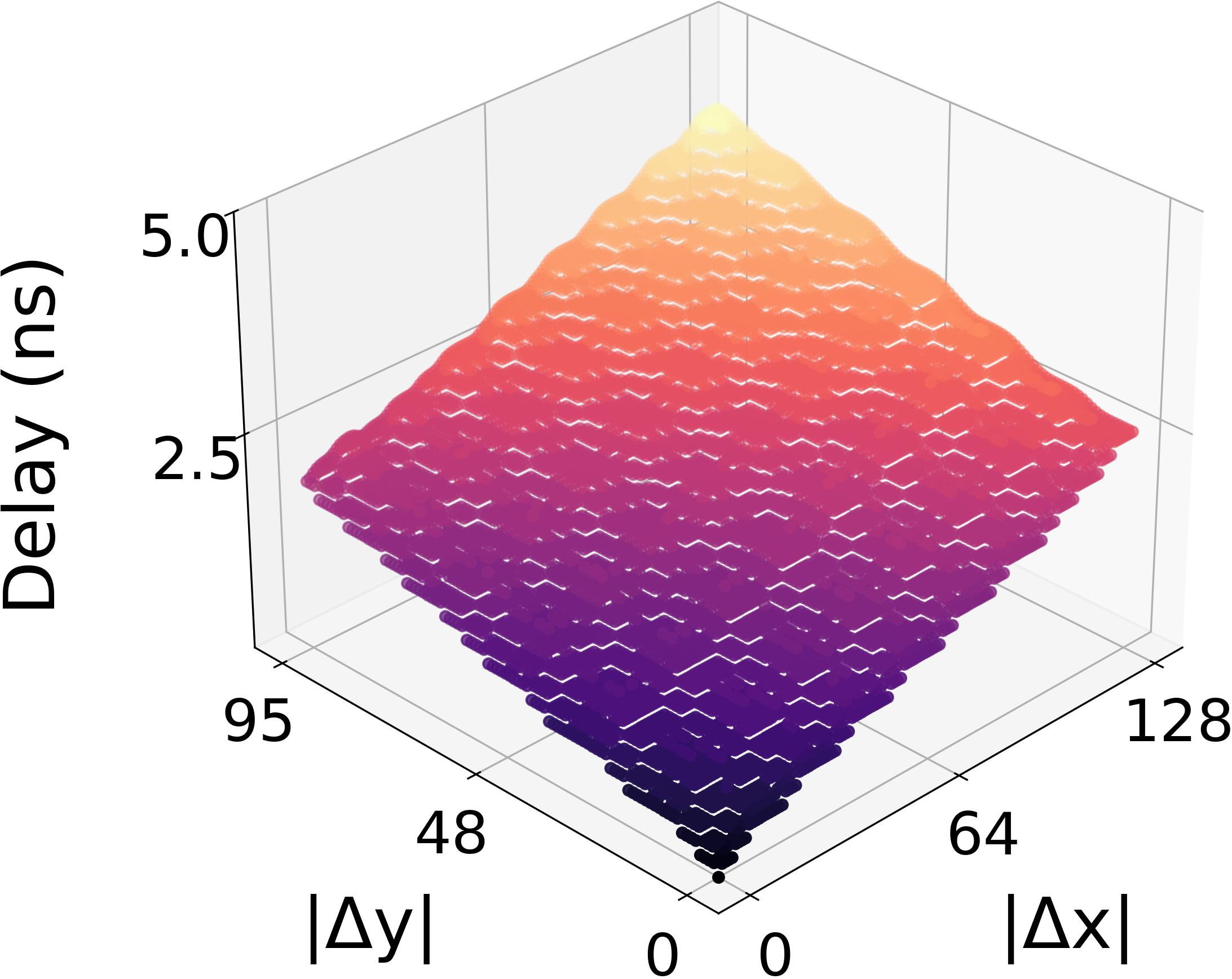}
         \caption{}
         \label{fig:fig:lookahead_int_zero}
     \end{subfigure}
     
     \caption{(a) Default router lookahead for a 2D architecture. (b) Default router lookahead for the same architecture but with an interposer cut in the middle. (c) Our modified router lookahead for the 2.5D architecture.}
     \label{fig:lookahead}
\end{figure}

VPR uses an $A^*$-based router that requires a heuristic cost estimate from each RR graph node to the sink. Recent VPR versions~\cite{vtr9} compute this estimate via a \emph{lookahead table}~\cite{air}: for each wire type and each $(|\Delta x|, |\Delta y|)$ offset, the table stores the minimum expected resource cost and delay to reach a node that far away. The table is filled by sampling routes from many starting wires to many target distances. For a 2D device with a uniform routing fabric, the resulting table is \emph{translation-invariant}: the estimated cost of moving $(|\Delta x|, |\Delta y|)$ is independent of the starting position, as illustrated in Fig.~\ref{fig:lookahead_no_int}.

In contrast, Nasiri et al.~\cite{tvlsi16} used a simpler, formula-based lookahead suited to routing architectures with a single wire type: it computed the $(x,y)$ distance from the current node to the sink, divided each component by the wire's length, and multiplied by the average delay per wire traversal. A fixed die-crossing penalty was added when the current node and sink resided on different dies, yielding a 31$\times$ routing speedup. This approach is insufficient for modern architectures, which mix short, medium, and long wire segments with complex inter-wire connectivity patterns, necessitating the table-based lookahead described above.

VPR's lookahead table cannot be used directly with 2.5D architectures either, however. Sampling does not yield a translation-invariant table in 2.5D devices because interposer wires introduce additional delay whenever a sample route crosses a die boundary, causing some table entries to be inflated. The result is a visible discontinuity at the interposer cut, as shown in Fig.~\ref{fig:fig:lookahead_int} for a device with an interposer cut at $y = 48$. This breaks translation invariance and degrades both routing quality and speed.

We solve this problem by decoupling the heuristic into two components: a \emph{translation-invariant intra-die lookahead} and a \emph{die-crossing cost table}. To build the intra-die lookahead, we modify the sampling procedure by setting the cost of all interposer wires to zero during sampling, preventing die-crossing delays from contaminating the table. As shown in Fig.~\ref{fig:fig:lookahead_int_zero}, the resulting table closely matches the regular structure of a 2D lookahead. For the die-crossing component, we compute, for each interposer cut, the minimum delay among all inter-die wire types that cross it, and store these minima in the die-crossing cost table. During routing, we add the die-crossing penalty for each cut that lies between the current node and the sink to the intra-die lookahead value. This decomposition resulted in a 2.5$\times$ reduction in routing time vs. the default VPR lookahead with no reduction in result quality.

\subsubsection{Placement Cost Function}
VPR's placer uses simulated annealing to minimize a  weighted sum of timing and wirelength costs. Wirelength is approximated by the half-perimeter wirelength (HPWL) of each net's bounding box. The timing cost is computed using a placement delay model, which is tied to the router lookahead; making the lookahead multi-die aware therefore makes the timing cost multi-die aware as well.

For interposer-based architectures, we include a \emph{cut} cost as in prior work on silicon-interposer FPGAs~\cite{hahn2014cad,tvlsi16}. Each net is penalized when its bounding box crosses an inter-die boundary, with the penalty growing linearly with the span of that box perpendicular to the boundary. In our early experiments, this term reduced the total number of nets crossing a die boundary, but did not encode \emph{where} along the boundary demand is concentrated. We still observed localized hotspots that caused routing failures. Ravishankar et al.~\cite{xil_place_strat} report the same issue and address it by inserting placeable inter-die wire driver instances into the netlist for each die-crossing net; the partition-driven placer then treats inter-die wires as capacity bins, guaranteeing no inter-die channel is oversubscribed. Instead of restructuring our annealing-based flow around a partition-driven approach, we adopt the congestion-aware placement cost of~\cite{shahrouz2025congestion}, which estimates channel demand via a wirelength per area model and penalizes nets whose average demand-to-capacity ratio within their bounding box exceeds a threshold. We restrict this congestion term to inter-die routing channels only, so the annealer directly discourages hotspot formation at die boundaries without interfering with intra-die placement.

\section{Experimental Results}\label{sec:experimental}
This section evaluates the impact of key inter-die routing architectural parameters including the fan-in/fan-out ($F_g$ and $F_{sc}$) of inter-die connections and the number of such connections per logic block ($N_{sg}$) on routability, critical path delay (CPD), routed wirelength (WL), and area in 3D/2.5D multi-die 7nm FPGAs. 

Each die uses the 7nm FPGA architecture from~\cite{into_third_dim} (10 fracturable 6-LUTs per LB, $W=300$, length-4/16 wires). Inter-die connectivity is modeled using the SG syntax from Section~\ref{sec:sg_syntax}, parameterized by varying $N_{sg}$, fan-in/fan-out ($F_{g}/F_{sc}$), and delays from Section~\ref{sec:detailed_modeling}. Benchmarks are drawn from the Koios benchmark suite~\cite{koios2}. For each circuit, the FPGA fabric is sized to the smallest square grid that fits the circuit, stressing inter-die routing resources.

As shown in Section~\ref{sec:int_feas}, sufficient 2.5D inter-die connectivity requires a minimum interposer wire length $L_{int}$, which sets a minimum device size of $2L_{int}\times2L_{int}$ tiles. Our maximum evaluated $L_{int}=37\,H_{LB}$ excludes circuits smaller than this threshold; the Koios subset we use spans 12k to 759k primitives and maps to $78\times78$ to $335\times335$ tile devices.

\subsection{2.5D Homogeneous Integration}

\begin{figure}
    \centering
    \includegraphics[width=0.65\linewidth]{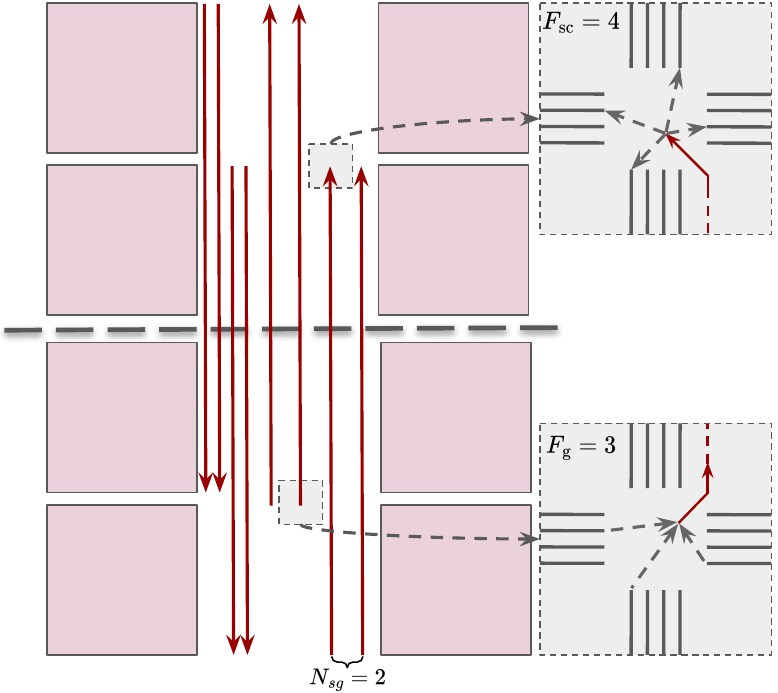}
    \caption{2.5D FPGA routing architecture ($N_{sg}=2$, $F_g=3$, $F_{sc}=4$, $L_{\text{int}}=3$). Zoomed-in sections show fan-in/fan-out between a single interposer wire and a switch-block location.}
    \label{fig:25d_routing_arch}
\end{figure}

In this section, we evaluate a family of 2.5D architectures consisting of two dies placed side by side, separated by a horizontal interposer cut that creates a discontinuity along the column direction; Fig.~\ref{fig:25d_routing_arch} illustrates an example 2.5D routing architecture. Based on the feasibility study of Section~\ref{sec:int_feas}, we evaluate pitch sizes of 10\,\textmu m and 5\,\textmu m for inter-die connections, with varying interposer wire lengths and driver fan-in. Table~\ref{tab:25d_res} summarizes the results of our 2.5D architecture exploration.

\begin{table}[t]
\centering
\begingroup
\newcommand{\PitchNsg}[2]{\makebox[2.15em][r]{#1}\,(#2)}%
\resizebox{\columnwidth}{!}{
\begin{tabular}{|c|c|c|c|c|c|c|c|}
\hline
\textbf{Pitch} & \multirow{2}{*}{\textbf{F\textsubscript{g}/F\textsubscript{sc}}} & \textbf{L\textsubscript{int}} & \multirow{2}{*}{\textalpha} & \multirow{2}{*}{\textbf{WL}} & \multirow{2}{*}{\textbf{CPD}} & \textbf{Route} & \textbf{LB} \\
\textbf{(N\textsubscript{sg})} & & \textbf{(LBs)} & & & & \textbf{Time} & \textbf{Area} \\ \hline \hline

2D & N.A. & N.A. & N.A. & 1 & 1 & 1 & 1 \\ \hline \hline

\multirow{9}{*}{\shortstack{10\,$\mu$m \\ (3) \\ \textsection~\ref{sec:lint}}} 
  & 48/64 & 5  & 0.08 & D.N.F. & D.N.F. & D.N.F. & 1.041 \\
  & 48/64 & 9  & 0.16 & D.N.F. & D.N.F. & D.N.F. & 1.042 \\
  & 48/64 & 13 & 0.24 & D.N.F. & D.N.F. & D.N.F. & 1.041 \\
  & 48/64 & 17 & 0.32 & 1.07   & 1.05   & 3.02   & 1.041 \\
  & 48/64 & 21 & 0.40 & 1.07   & 1.06   & 5.73   & 1.042 \\
  & 48/64 & 25 & 0.48 & 1.10   & 1.07   & 10.00  & 1.041 \\
  & 48/64 & 29 & 0.56 & 1.14   & 1.08   & 18.30  & 1.041 \\
  & 48/64 & 33 & 0.64 & 1.19   & 1.10   & 32.28  & 1.036 \\
  & 48/64 & 37 & 0.72 & 1.21   & 1.11   & 48.09  & 1.036 \\ \hline \hline

\multirow{5}{*}{\shortstack{10\,$\mu$m \\ (3) \\ \textsection~\ref{sec:fanin}}} 
  & 3/4   & 17 & 0.32 & D.N.F. & D.N.F. & D.N.F. & 1.016 \\
  & 6/8   & 17 & 0.32 & 1.12   & 1.06   & 4.12   & 1.021 \\
  & 12/16 & 17 & 0.32 & 1.07   & 1.05   & 3.28   & 1.024 \\
  & 24/32 & 17 & 0.32 & 1.06   & 1.05   & 3.09   & 1.029 \\
  & 48/64 & 17 & 0.32 & 1.07   & 1.05   & 3.04   & 1.041 \\ \hline \hline

\PitchNsg{}{12} & 12/16 & 5  & 0.32 & D.N.F. & D.N.F. & D.N.F. & 1.048 \\
\PitchNsg{}{6}  & 12/16 & 9  & 0.32 & 1.02   & 1.04   & 1.62   & 1.025 \\
\PitchNsg{5\,$\mu$m}{4} & 12/16 & 13 & 0.32 & 1.04   & 1.04   & 2.04   & 1.012 \\
\PitchNsg{\textsection~\ref{sec:density}}{3} & 12/16 & 17 & 0.32 & 1.06   & 1.05   & 3.18   & 1.008 \\
\PitchNsg{}{2}  & 12/16 & 25 & 0.32 & 1.11   & 1.08   & 9.81   & 1.004 \\ \hline

\end{tabular}
}
\endgroup
\caption{Comprehensive study of interposer wire lengths, $F_g/F_{sc}$ ratios, and pitch configurations. All results are the geometric mean of the benchmark suite normalized to the 2D baseline. Entries denoted as D.N.F. (did not finish) indicate architectures that were not able to route all benchmark circuits.}
\label{tab:25d_res}
\end{table}

\subsubsection{Interposer Wire Length}\label{sec:lint}

For a fixed bump pitch, increasing interposer wire length ($L_{int}$) raises the inter-die connectivity ratio $\alpha$ at the cost of higher wire delay. Using an architecture with a bump pitch of 10\,\textmu m and interposer wire fan-in of 48, we test $L_{int} \in \{5, 9, 13, 17, 21, 25, 29, 33, 37\}$. The shortest $L_{int}$ that achieves routing success across all benchmarks is $L_{int} = 17$, corresponding to $\alpha = 0.32$. This is consistent with results from~\cite{tvlsi16} showing that routability is not severely impacted when $\alpha > 0.25$. For routable architectures, increasing $L_{int}$ improves $\alpha$ but directly degrades CPD due to higher wire delay. Beyond CPD, longer $L_{int}$ also increases WL and route time, which we attribute to the behavior of placement and routing algorithms. The placement cost function minimizes HPWL by placing connected netlist primitives close together; however, when two connected blocks are on opposite sides of an interposer cut, placing them near the boundary does not help, as they cannot be connected through short local wires and must instead use the long interposer wire spanning $L_{int}$ tiles. On the routing side, VPR's router uses the bounding-box heuristic~\cite{air} to prune the RR graph search to a slightly expanded bounding box of each net. Long interposer wires can extend beyond this bounding box, causing search failures that force the router to repeatedly enlarge it and increase runtime. These observations highlight that placement and routing algorithms have significant room for improvement for 2.5D devices, beyond the multi-die aware enhancements introduced in Section~\ref{sec:cad_flow}.

\subsubsection{Interposer Wire Fan-in}\label{sec:fanin}

Since increasing $L_{int}$ beyond the minimum required for reliable routability only worsens CPD, we select $L_{int} = 17$ and sweep the interposer wire fan-in, testing $F_{g} \in \{48, 24, 12, 6, 3\}$. In this architecture, the fan-in wires are gathered from the sides of each switch block that face away from the interposer cut, and $F_{sc}$ is set to $\frac{4}{3} F_{g}$ at the destination.

At $F_{g} = 3$, each inter-die connection aggregates only 3 intra-die wires, making it difficult for the router to find an interposer wire to cross dice with, resulting in unroutable circuits. Increasing $F_{g}$ improves routability and reduces route time, as wider muxes make inter-die routing resources more accessible.
Area overhead grows substantially with fan-in: increasing $F_{g}$ from 3 to 48 more than doubles the LB area overhead due to significant multiplexer area. CPD and WL both decrease with larger $F_{g}$, as wider muxes reduce intra-die routing detours, though gains diminish beyond $F_{g} = 12$. Fig. \ref{fig:ad_fig} shows how the area-delay product of a 2.5D architecture with 10 {\textmu}m bumps and an interposer wire length of 17 tiles varies with $F_{g}$. 
The best values of $F_{g}$ are in the 8 to 32 range, which balance routability with the area and delay of the added mux. 

\subsubsection{Inter-die Connection Density}\label{sec:density}

As shown in Table~\ref{tab:int_wl_vs_crossing}, achieving full inter-die connectivity ($\alpha = 1$) is feasible with fine-pitch hybrid bonding. For a $5\,\mu\text{m}$ pitch, $\alpha = 1$ requires $L_{int} = 13$, meaning that only 12 LB rows on each side of the boundary need to accommodate the interposer wire circuitry. However, since FPGAs are manufactured by tiling the same LB across the entire device, a column structure must be maintained: even if only a fraction of the tiles in a column require wider layout to host the additional driving circuitry and ESD protection, all tiles in that column must match that width. As a result, the area overhead extends to the full column height, not just the rows directly involved in inter-die connectivity.

This observation motivates using longer interposer wires to spread the interposer wire circuitry across a larger number of LB rows, reducing the per-tile area overhead by scattering buffers, drivers, and ESD circuitry over a larger region rather than concentrating them near the die boundary. To evaluate this trade-off, we fix the $5\,\mu\text{m}$ pitch and hold $\alpha = 0.32$ while increasing $L_{int}$ from 5 to 25. The results are shown in the bottom section of Table~\ref{tab:25d_res}. Note that $L_{int} = 5$ itself is not reliably routable: inter-die wires must be reached through intra-die routing resources, and concentrating all of them in a narrow band immediately adjacent to the boundary creates local routing congestion in that region. Longer interposer wires distribute this demand over a wider area, alleviating the congestion. Tile area decreases by 4.2\% across this range, confirming that spreading the circuitry over more rows meaningfully reduces per-tile overhead. Delay increases for interposer wires longer than 13 LBs due to their higher wire delay, and the best area-delay tradeoff occurs with $L_{int}$ in the 13 to 17 range. As previously discussed, route time increases with long interposer wires, motivating future work in interposer-aware routing algorithms.

\subsection{3D-stacked Homogeneous Integration}
The 3D FPGA architecture evaluated in this study comprises two homogeneous dice stacked atop each other. For inter-layer connectivity, we consider three die stacking technologies: $1\,\mu\text{m}$ and $5\,\mu\text{m}$ pitch hybrid bonds, and $10\,\mu\text{m}$ $\mu$bumps. There are multiple possible routing architectures to leverage these inter-die links. One option is to cross between dice only at output pins, as in~\cite{into_third_dim}; we call an architecture where all block output pins can reach the CHANZ nodes that model die-crossing a \textit{full output pin connectivity} architecture. Another possibility is to have routing wires on one layer drive the bumps (CHANZ nodes) that reach routing wires on a different die layer; we call this option \textit{switch-block connectivity}. The choice between full output pin or switch-block connectivity depends on the available inter-layer connection density. For the $5\,\mu\text{m}$ and $10\,\mu\text{m}$ pitches, where the inter-layer connection density is insufficient to support full output pin connectivity, the inter-layer interface is realized by $N_{sg}$ vertical connections per direction; here, each vertical connection is driven by a multiplexer aggregating $F_{g}$ length-4 wires from all four cardinal sides of the source switch block. Driving CHANZ nodes from both x- and y-directed routing wires via the gather multiplexer maximizes use of the scarce die-crossing bumps, as each bump can be driven by any of several candidate wires and signals can cross dice anywhere along a net's path. At the $1\,\mu\text{m}$ pitch, the inter-layer connection density suffices for full output pin connectivity: LB output pins drive CHANZ nodes directly ($F_g=1$) without a gather multiplexer. In both full output pin and switch-block connectivity, upon reaching the adjacent die, the inter-layer signal fans out into $F_{sc}$ length-4 wires, which are distributed across the four sides of the destination switch block. Length-16 wires and input pins do not connect to the inter-layer interconnects directly.

For the $5\,\mu\text{m}$ and $10\,\mu\text{m}$ pitches, we integrate the maximum possible number of vertical resources, yielding $N_{sg}$ values of 12 and 3 connections per direction per LB, respectively. For the $1\,\mu\text{m}$ pitch, rather than maximizing the number of connections (which would yield up to $N_{sg}=302$), we set $N_{sg}=40$ so that every output pin can cross to the other die via its own dedicated inter-layer connection. To investigate the threshold of routability and the impact of inter-layer routing resource scarcity, we further evaluate a constrained $10\,\mu\text{m}$ case where $N_{sg}$ is reduced to 1. Furthermore, we modeled inter-layer connectivity with a $25\,\mu\text{m}$ $\mu$bump pitch, which translates to a density of $N_{sg}=0.5$ connections per direction per LB. In this configuration, vertical connections are staggered such that the connection direction alternates at each LB. More than one-third of the benchmark circuits failed to route with a $25\,\mu\text{m}$ pitch, suggesting that this inter-layer connection density falls well below the threshold of routability. Across the remaining viable combinations of pitch and $N_{sg}$, we sweep inter-layer connection fan-in and fan-out, testing $F_g, F_{sc} \in \{64, 32, 16, 8, 4\}$. Results are reported in Table~\ref{tab:3d_qor_impact_updated}.

\begin{table}[t]
\centering
\resizebox{\columnwidth}{!}{
\begin{tabular}{|c|c|c|c|c|c|c|c|}
\hline
\textbf{Pitch} & \multirow{2}{*}{\textbf{F\textsubscript{g}/F\textsubscript{sc}}} & \textbf{3D} & \multirow{2}{*}{\textbf{WL}} & \multirow{2}{*}{\textbf{CPD}} & \textbf{Route} & \textbf{LB} & \textbf{ESD} \\
\textbf{(N\textsubscript{sg})} & & \textbf{Nets} & & & \textbf{Time} & \textbf{Area} & \textbf{Area} \\ \hline \hline

2D & N.A. & N.A. & 1.00 & 1.00 & 1.00 & N.A. & N.A. \\ \hline \hline

1 $\mu$m (40) & 1/64 & 1.00 & 0.90 & 0.95 & 0.92 & 1.104 & 6.29\% \\ \hline \hline

\multirow{5}{*}{\shortstack[c]{5 $\mu$m \\ (12)}}
  & 64/64 & 0.84 & 0.86 & 0.96 & 1.07 & 1.091 & 1.91\% \\
  & 32/32 & 0.83 & 0.87 & 0.96 & 0.99 & 1.068 & 1.95\% \\
  & 16/16 & 0.82 & 0.90 & 0.96 & 1.00 & 1.051 & 1.98\% \\
  & 8/8  & 0.82 & 0.92 & 0.95 & 1.04 & 1.043 & 2.00\% \\
  & 4/4  & 0.83 & 0.93 & 0.96 & 1.11 & 1.034 & 2.01\% \\ \hline \hline

\multirow{5}{*}{\shortstack[c]{10 $\mu$m \\ (3)}}
  & 64/64 & 0.61 & 0.89 & 0.96 & 0.92 & 1.023 & 0.51\% \\
  & 32/32 & 0.61 & 0.90 & 0.95 & 0.94 & 1.017 & 0.51\% \\
  & 16/16 & 0.60 & 0.92 & 0.95 & 0.98 & 1.013 & 0.51\% \\
  & 8/8  & 0.61 & 0.93 & 0.96 & 1.01 & 1.011 & 0.51\% \\
  & 4/4  & 0.61 & 0.94 & 0.97 & 1.07 & 1.008 & 0.52\% \\ \hline \hline

\multirow{5}{*}{\shortstack[c]{10 $\mu$m \\ (1)}}
  & 64/64 & 0.54 & 0.91 & 0.95 & 1.00 & 1.008 & 0.17\% \\
  & 32/32 & 0.55 & 0.92 & 0.94 & 1.00 & 1.006 & 0.17\% \\
  & 16/16 & 0.54 & 0.93 & 0.95 & 1.07 & 1.004 & 0.17\% \\
  & 8/8  & 0.54 & 0.95 & 0.98 & 1.11 & 1.004 & 0.17\% \\
  & 4/4  & 0.54 & 0.96 & 0.98 & 1.19 & 1.003 & 0.17\% \\ \hline

\end{tabular}
}
\caption{\small{Impact of pitch, $N_{sg}$, $F_g$ and $F_{sc}$ on QoR and area. All QoR metrics are normalized to the 2D baseline. LB Area is reported with ESD included; the ESD Area column lists the percentage of LB area due to ESD.}}
\label{tab:3d_qor_impact_updated}
\end{table}

\subsubsection{Routability under Scarce Inter-layer Connectivity}
A key finding is that 3D homogeneous integration does not require the most advanced and expensive packaging technologies to achieve routability. Prior work assumed high inter-layer connectivity: Boutros et al.~\cite{into_third_dim} used full output-pin connectivity ($N_{sg}=40$), and Lazagna~\cite{youssef2025lazagna} demonstrated routability with $N_{sg} \approx 26$ using 3D switch blocks. In both cases, the required connection density can only be realized with a $1\,\mu\text{m}$ hybrid-bond pitch. Our results show that circuits remain routable down to $N_{sg}=1$, which is achievable with $10\,\mu\text{m}$ $\mu$bumps, a more cost-effective packaging technology. Beyond reducing manufacturing cost, lower inter-layer connectivity also shrinks the per-tile area overhead of the multiplexers, buffers, and ESD protection needed to implement the inter-layer interface, further improving the overall area efficiency of the 3D FPGA fabric. Achieving routability under scarce inter-layer connectivity, however, does not depend on interconnect density alone; placement must also be aware of how limited these resources are. This is handled through the HPWL cost, which scales displacement along each dimension by the inverse of available routing resources~\cite{mohaghegh2024expanding}. As $N_{sg}$ decreases, the relative cost of crossing to another die rises, encouraging the placer to keep net endpoints within a single die. This is reflected in the results: the number of inter-layer nets is sensitive to $N_{sg}$ but largely independent of $F_{g}$ and $F_{sc}$, since $F_{g}$ and $F_{sc}$ affect routing flexibility rather than resource density.

\subsubsection{QoR and Route Time Impact}
Reduced inter-layer connectivity imposes a measurable cost on both QoR and route time. As $N_{sg}$ decreases, the router faces a scarcer pool of die-crossing routing resources, increasing the likelihood of congestion at die boundaries. When local inter-layer connections are occupied, the router must detour to find available CHANZ nodes further away, driving up both WL and route time. CPD is also affected, though less severely, as the timing-driven router reserves local die-crossing connections for timing-critical nets and reroutes non-critical nets along longer paths. Inter-layer connection density alone, however, does not fully determine QoR, as the accessibility of those connections matters equally. As $F_g$ or $F_{sc}$ decreases, each inter-layer connection aggregates fewer intra-die wires, making it harder for the router to find a suitable die-crossing resource for a given net. This reduced accessibility leads to longer routing detours, increasing WL, CPD, and route time even when the total number of die-crossing connections remains fixed. The most accessible configuration is full output-pin connectivity ($1\,\mu\text{m}$ pitch, $F_g=1$), where every output pin has its own dedicated inter-layer wire and can immediately cross to the adjacent die without competing for a shared multiplexer. This yields the best route time and CPD among all evaluated architectures, with 23\% faster routing and 3\% lower CPD compared to the most resource-scarce configuration ($10\,\mu\text{m}$, $N_{sg}=1$, $F_g = F_{sc}=4$).

\subsubsection{Area-Delay Trade-off}
\begin{figure}[t]
    \centering
    \includegraphics[width=0.9\linewidth]{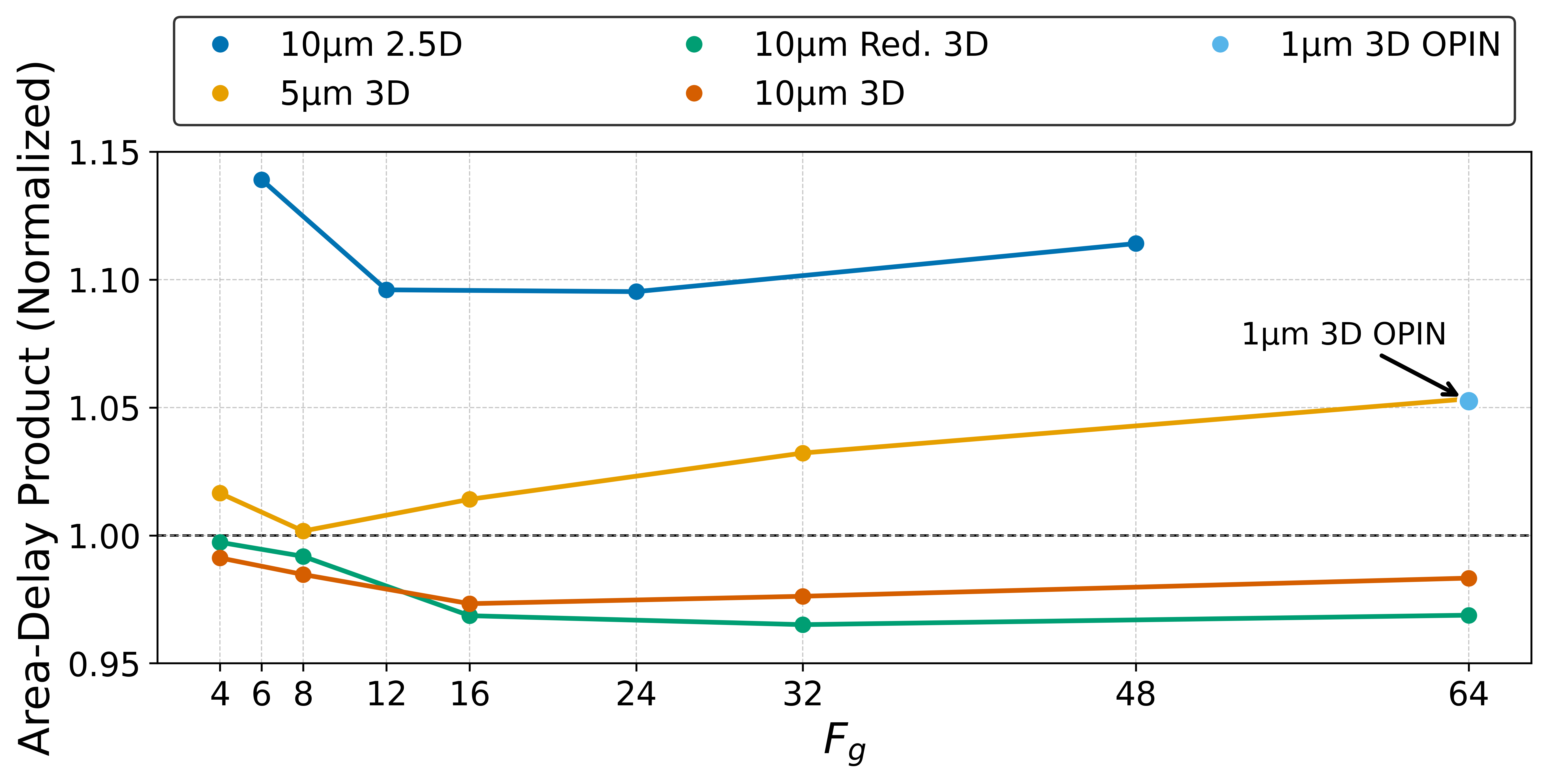}
    \caption{Area-delay products of the explored multi-die architectures.}
    \label{fig:ad_fig}
\end{figure}

Tile area moves in the opposite direction to WL and CPD, revealing a fundamental trade-off. Reducing $N_{sg}$ not only enables 3D integration with less advanced packaging technologies, as discussed above, but also lowers the silicon area overhead: fewer inter-layer connections mean fewer buffers, multiplexers, and ESD protection circuits per tile. Reducing $F_{g}$, which makes connections less accessible by narrowing the input multiplexers, has the same effect, as smaller muxes occupy less area. Thus, the architectural choices that improve QoR and reduce routing runtime (higher $N_{sg}$, larger $F_{g}$) come at the cost of increased tile area, while leaner configurations trade some QoR for a smaller silicon footprint. The area cost of larger $F_{g}$ is more pronounced in architectures with finer bump pitches and higher $N_{sg}$. Fig.~\ref{fig:ad_fig} shows the area-delay product of the evaluated 3D architectures. The $10\,\mu\text{m}$ architecture with $N_{sg}=1$ achieves the best area-delay trade-off among all evaluated architectures, meaning that the area savings in this architecture outweigh the resulting CPD degradation.

Recent packaging technology advancements have demonstrated that ESD protection circuitry can be simplified or even eliminated~\cite{lin2024toward,haque2025tiny}. As  Table~\ref{tab:3d_qor_impact_updated} shows, the area saved by removing ESD protection is most significant in architectures with a high density of inter-layer connections. For the $1\,\mu\text{m}$ pitch ($N_{sg}=40$), removing ESD yields an area saving of 6.29\%, whereas these savings diminish as vertical connections become sparser, reaching as low as 0.17\% for the $10\,\mu\text{m}$ pitch ($N_{sg}=1$) cases.

\section{Conclusion}

Advanced packaging enables higher-capacity multi-die FPGAs, but inter-die routing must be tailored to each technology's constraints, motivating versatile CAD modeling tools. To address this need, we presented detailed circuit models of inter-die connections in 2.5D and 3D FPGAs, enabling accurate delay and area estimates for a range of die-crossing technologies and fan-in/fan-out configurations. We also introduced a first-order feasibility analysis for 2.5D architectures that links bump pitch, interposer wire length, and inter-die connectivity ratio, enabling rapid pruning of the design space before detailed simulation. We also extended the VTR architecture description language with the scatter-gather (SG) syntax, modeling inter-die connections with independent control over fan-in, fan-out, wire length, and connectivity pattern.

Using these tools, we conducted a detailed design space exploration of 3D and 2.5D FPGA routing architectures. For 3D FPGAs, we show that homogeneous die stacking does not require the most advanced packaging technologies to achieve reliable routability. With $10\,\mu\text{m}$ $\mu$bump technology, 3D FPGAs remain routable even with as few as two inter-layer connections per logic block using a \textit{switch block connectivity} 3D architecture. High-density $1\,\mu\text{m}$ hybrid bonding enables a \textit{full output-pin connectivity} architecture, and 3D FPGAs achieve up to 10\% wirelength reduction and 5\% CPD improvement. For 2.5D FPGAs, we find that approximately 32\% inter-die connectivity is sufficient for reliable routability, incurring only a 2\% wirelength and 4\% CPD overhead relative to a 2D baseline. The inter-die architecture specification syntax and CAD optimizations introduced in this work are open source and integrated with the VTR master branch\footnote{\href{https://github.com/verilog-to-routing/vtr-verilog-to-routing}{github.com/verilog-to-routing/vtr-verilog-to-routing}}, enabling future enhancements and work in this area.

\section*{Acknowledgements}
The authors thank Altera, NSERC, and QuickLogic for research funding and Amin Mohaghegh for insightful technical discussions.

\bibliographystyle{IEEEtran}
\bibliography{references}{}

\end{document}